\documentclass{emulateapj}

\shorttitle{Identify SSs from GRB magnetars} \shortauthors{Yu et
al.}

\newcommand{\mn}{\mbox{MNRAS}}
\newcommand{\aaa}{\mbox{A\&A}}

\newcommand{\beq}{\begin{equation}}
\newcommand{\mes}{M\'esz\'aros}

\renewcommand{\r}{{\it r}-mode}

\def\be{\begin{equation}}
\def\ee{\end{equation}}

\begin{document}
\title{The newly born magnetars powering gamma-ray burst internal-plateau emission: are there strange stars?}

\author{Yun-Wei Yu\altaffilmark{1}}
\author{Xiao-Feng Cao\altaffilmark{1}}
\author{Xiao-Ping Zheng\altaffilmark{1}}

\altaffiltext{1}{Institute of Astrophysics, Huazhong Normal
University, Wuhan, 430079, China; yuyw@phy.ccnu.edu.cn}

\begin{abstract}
The internal-plateau X-ray emission of gamma-ray bursts (GRBs)
indicates that a newly born magnetar could be the central object of
some GRBs. The observed luminosity and duration of the plateaus
suggest that, for such a magnetar, a rapid spin with a sub- or
millisecond period is sometimes able to last thousands of seconds.
In this case, the conventional neutron star (NS) model for the
magnetar may be challenged, since the rapid spin of nascent NSs
would be remarkably decelerated within hundreds of seconds due to
{\r} instability. In contrast, the {\r}s can be effectively
suppressed in nascent strange stars (SSs). In other words, to a
certain extent, only SSs can keep nearly-constant extremely-rapid
spin for a long period of time during the early ages of the stars.
We thus propose that the sample of the GRB rapidly-spinning
magnetars can be used to test the SS hypothesis based on the
distinct spin limits of NSs and SSs.
\end{abstract}


\slugcomment{2009, ApJ, ??, ??}

\section{Introduction}
%Gamma-ray bursts (GRBs) are short and intense flashes of soft
%gamma-rays ($\sim0.01-1$ MeV), followed by long-lasting afterglow
%emission in the X-ray, optical, and radio wavelengthes.
In the standard fireball model for gamma-ray bursts (GRBs;
%Rees \& \mes 1994; \mes \& Rees 1997; Sari et al. 1998;
see reviews by Piran 2005; {\mes} 2005), the GRB prompt emission is
considered to arise from an internal dissipation in a relativistic
expanding fireball, while the afterglow emission is produced by the
deceleration of the fireball by circumburst medium, i.e., a blast
wave. As understood above, it seems unlikely to directly find
information about the GRB central engine, which drives the fireball,
from the observed emission. However, based on some observational
constraints and theoretical simulations, it is widely accepted that
black holes or rapidly spinning magnetars (highly-magnetized
pulsars) formed during the death of the progenitors play an
important role in driving the fireball.

%In the canonical X-ray light curve summarized from a great
%fraction of the \textit{Swift} GRBs (Nousek et al 2006; Zhang et
%al. 2006),
%a shallow decaying segment
%appears remarkably from a few hundred seconds to a few hours (even
%days), following a rapidly decaying segment and preceding a normal
%decaying segment . It is
%generally considered that this shallow decaying afterglow emission
%probably
%the appearance of the shallow decaying segment is generally
%considered to indicate an energy injection into the GRB blast
%waves (i.e., external shock). In view of the possible
%inhomogeneity of the GRB ejecta, the energy injection could be
%ascribed to the effect of the slower but more massive tail of the
%GRB ejecta (Rees \& {\mes} 1998). In this case, there is no
%connection between the shallow decaying afterglows and the
%postburst central objects. However,

%since the launch of the {\it Swift} spacecraft, some stricter
%constraints on the properties of the central objects have been
%given by the accumulated abundant refined X-ray light curves (see
%recent reviews by Zhang 2007; Gehrels et al. 2009).
Furthermore, the {\it Swift}-discovered delayed intermittent bright
X-ray flares demonstrate that the GRB central objects should be
still very active after the bursts (Burrows et al. 2005; Yu \& Dai
2009). Analogically, the shallow-decaying segment remarkably
appearing in the canonical X-ray light curve (Nousek et al 2006;
Zhang et al. 2006) also strongly implies an energy injection into
the GRB blast waves, which is quite likely to result from a
long-lasting energy release from the central objects. Under these
requirements, spinning-down magnetars are usually suggested as GRB
central objects in the literature (e.g., Bucciantini et al. 2009;
Corsi \& {\mes} 2009; Dai 2004; Dai et al. 2006; De Pasquale et al.
2007; Zhang \& Dai 2008, 2009). The shallow-decaying afterglows can
somewhat be regarded as an observational signature of rapidly
spinning magntetars (Dai \& Lu 1998a, b; Yu \& Dai 2007; Zhang \&
M\'esz\'aros 2001).

In particular, a nearly-constant X-ray emission followed by a very
steep decline with a temporal index of $\sim9$ was reported in GRB
070110 by Troja et al. (2007). The abrupt cutoff of the X-ray
light curve robustly indicates that this plateau emission is of
``internal" origin, ruling out the external shock model. Moreover,
the observed luminosity and duration of the plateau are well
consistent with the parameters of a newly born magnetar. So,
following Zhang \& {\mes} (2001), Troja et al. (2007) further
argued that this puzzling plateau could be directly produced by
the internal dissipation of the magnetar-driven wind at small
radii, before the energy of the wind is deposited into the GRB
blast wave.
%In this
%case, the abrupt cutoff might reflect a rapid decrease of the
%stellar magnetic field (Troja et al. 2007) or a collapse of the
%magnetar into a black hole (Lyons et al. 2009).
Then, the observed luminosity of the plateau tracking the
spin-down luminosity of the GRB magnetar makes it possible to
``directly" explore the properties of the central objects by using
observed emission.

As a pioneering attempt, Lyons et al. (2009) investigated the
magnetic and rotational properties of the central magnetars for ten
GRB 070110-like GRBs. While the inferred magnetic field strengthes
are basically typical for magnetars, the spin periods are usually
found to be as short as the Kepler period, sometimes even at
thousands of seconds after the GRB trigger. However, for a magnetar
composing of normal neutron matter (i.e., neutron star; NS), it
seems difficult to keep nearly-Kepler spin for more than one
thousand seconds, because the spin of the nascent NS could be sorely
restricted by some stellar instabilities, especially, {\r}
instability on focus in this {\it letter} (S\'a 2004; S\'a \& Tom\'e
2005, 2006; Yu et al. 2009). In contrast, for a magnetar composing
of strange quark matter (i.e., strange star; SS)\footnote{According
to the different physics of dense matter, a variety of models for
compact stars have been constructed in the past forty years (see
reviews by Weber 2005; Weber et al. 2006). Most of these models can
be regarded as advanced versions of the conventional NS model.
However, as an extreme case, SSs composing of strange quark matter
(usually with a thin nuclear crust) are predicted to be completely
different from ordinary NSs (e.g., Alcock et al. 1986).}, the \r
s-induced limit on the stellar spin is absent during the early ages
of the star, because of the effective suppression of the {\r}s by
the quark matter's bulk viscosity (e.g., Wang \& Lu 1984; Madsen
1998; Zheng et al. 2006). Therefore, we argue that {\it a newly born
magnetar, which can maintain constant rapid spin for a long period
of time, could be a SS candidate}. Based on this consideration, we
propose that {\it the sample of the GRB rapidly-spinning magnetars
can be used to test the SS hypothesis.}

In the next section, we briefly review the magnetar model for the
internal-plateau emission, and then some magnetar parameters are
derived from a small GRB sample. In Sect. 3, we analysis the {\r}
limit on the spin of NSs in the framework of a second-order {\r}
model developed by S\'a (2004). By confronting the {\r} limit with
the GRB magnetar sample, we makes an attempt to find SS
candidates. Finally, a summary and discussion are given in Sect.
4.
\section{Magnetars powering internal plateaus}
As discussed above, the internal-plateau emission indicates that a
rapidly spinning magnetar has been formed as a central object during
the prompt phase of some GRBs. The initial spin-down of a magnetar
could be very complicated during the first tens of seconds after its
birth. On the longer timescales that we are interested in, some
short-term processes such as neutrino-driven winds could be no
longer important. We here simply consider that the nascent magnetar
spins down through magnetic dipole radiation as
\begin{eqnarray}
{dP\over dt}={P\over \tau_m},\label{magbrake}
\end{eqnarray}
where $P$ is the spin period of the star. The magnetic braking
timescale reads (Shapiro \& Teukolsky 1983)
\begin{eqnarray}
\tau_m&=&{6c^3\over(2\pi)^2}{IP^2\over
B^2R^6}=4\times10^3I_{45}R_6^{-6}B_{15}^{-2}P_{-3}^{2}~\rm s,
\end{eqnarray}
where $c$ is the speed of light, $I$, $R$, and $B$ are the
inertial of moment, the radius, and the surface magnetic field
strength of the magnetar, respectively. The convention
$Q_x=Q/10^x$ is adopted in cgs units hereafter. Eq.
(\ref{magbrake}) yields $P(t)=P_i(1+t/T_{m})^{1/2}$, where $P_i$
is the initial spin period and
\begin{eqnarray}
T_{m}={P_i^2\over2P^2}\tau_m={1\over2}\tau_{m,i}=2\times10^3I_{45}R_6^{-6}B_{15}^{-2}P_{i,-3}^{2}~\rm
s.
\end{eqnarray}
The expression of $P(t)$ shows that a nascent magnetar only against
the magnetic braking effect can keep nearly-constant spin for a
period of time of $T_{m}$. As the spinning down, the magnetar
releases the spining energy and drives an outwards-propagating wind.
We then approximately estimate the luminosity of the wind by the
magnetically spin-down luminosity as $L_w\approx
L_{msd}={(2\pi)^2I\dot{P}/ P^3}=\tilde{L}(1+t/T_{m})^{-2}$ with
\begin{eqnarray}
\tilde{L}\equiv{(2\pi)^4\over6c^3}{ B^2R^6\over
P_i^4}=10^{49}R_6^{6}B_{15}^{2}P_{i,-3}^{-4}~\rm erg~s^{-1}.
\end{eqnarray}

Since the observed internal-plateau emission is deemed to be
directly emitted by magnetar winds, the model quantities
$\tilde{L}$ and $T_{m}$ can be easily determined by the
observational isotropically-equivalent luminosity $L_{\gamma,iso}$
and the observational duration $T_{p}^{obs}$ of the plateaus as
follows
\begin{eqnarray}
\tilde{L}\geq L_{\gamma,b}=f_{b}L_{\gamma,iso},~~~T_{m}\geq
T_{p}={T_{p}^{obs}\over(1+z)}\label{L and tau},
\end{eqnarray}
where $z$ is the redshifts of the GRBs. The beaming factor
$f_b=(1-\cos\theta_w)$ with $\theta_w$ being the half-opening angle
of the winds is introduced because the magnetar winds could be
collimated rather than perfectly isotropic (Bucciantini et al. 2007;
Lyons et al. 2009). By solving Eq.(\ref{L and tau}), the magnetic
field strengthes and the initial spin periods of the GRB magnetars
can be derived as
\begin{eqnarray}
B&\leq&2\times10^{15}I_{45}R_6^{-3}L_{\gamma,b,49}^{-1/2}T_{p,3}^{-1}~\rm
Gauss,\label{B}\\
P_{i}&\leq&1.4\times10^{-3}I_{45}^{1/2}L_{\gamma,b,49}^{-1/2}T_{p,3}^{-1/2}~{\rm
s}\label{Pi}.
\end{eqnarray}
The greater-than and less-than signs appear in Eqs. (\ref{L and
tau})-(\ref{Pi}), because the radiation efficiency of the winds
should be lower than 100\% and the cutoff of the plateaus could be
caused by a rapid decrease of the stellar magnetic fields (Troja et
al. 2007) or by a collapse of the magnetars into black holes (Lyons
et al. 2009). In this {\it letter}, for simplicity, we only consider
the case of equality for Eqs. (\ref{B}) and (\ref{Pi}) as an extreme
case, as did in Lyons et al. (2009).

\begin{figure}
\plotone{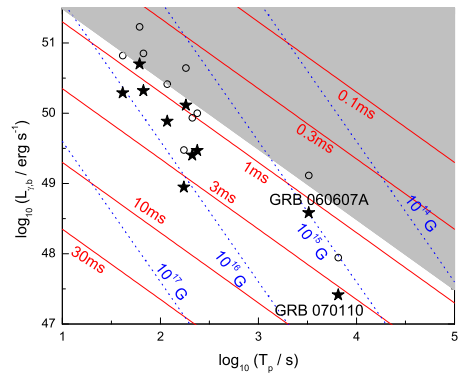} \caption{Equal-$B$ (dotted lines) and equal-$P_i$
(solid lines) contours determined by Eqs. (\ref{logLTB1}) and
(\ref{logLTP1}), respectively. The observational internal-plateau
data of ten GRBs are marked by the open circles for isotropic winds
and solid asterisks for $\theta_{w}=45^{\circ}$. The shaded region
is forbidden for magnetars due to $P_i<P_K$.} \label{figure:2}
\end{figure}

\begin{table}
\caption{The observed properties of the internal plateaus of ten
GRBs$^*$.}
\begin{tabular}{ c c c c c }
\hline
GRB & $L_{\gamma,iso}$ (erg s$^{-1}$) & $T_p^{obs}$ (s) & $z$ & $T_p$ (s) \\
\hline
 080310     & 2.6e+50 & 401.9  & 2.426       &  117.3  \\
 071021     & 6.6e+50 & 248.3  & 5.0         &  41.4\\
 070721B    & 3.0e+49 & 802.9  & 3.626       &  173.6 \\
 070616     & 4.4e+50 & 585.6  & 2.22        &  181.9\\
 070129     & 8.6e+49 & 683.0  & 2.22        &  212.1 \\
 070110     & 8.8e+47 & 21887.1& 2.352       &  6529.6 \\
 060607A    & 1.3e+49 & 13294.7 & 3.082      &  3256.9\\
 060510B    & 1.7e+51 & 362.9  & 4.9         &  61.5 \\
 060202     & 1.0e+50 & 766.0  & 2.22        &  237.9\\
 050904     & 7.1e+50 & 488.8  & 6.29        &  67.1 \\
\hline
\end{tabular}
\\
$^*$See Lyons et al. (2009) for a detailed explanation of the
data.
\end{table}

In order to conveniently estimate the values of $B$ and $P_i$ for
specific GRBs in the $\lg L_{\gamma,b}-\lg T_p$ panel, we plot in
Fig. 1 a set of dashed lines for different values of $B$ and a set
of dotted lines for different values of $P_i$, which are
respectively determined by
%\begin{eqnarray}
%\lg L_{\gamma,b}&=&-2\lg T_{p}+\left(2\lg I_{45}-6\lg R_6-2\lg
%B_{15}-\lg
%f_b+55.6\right),\label{logLTB1}\\
%\lg L_{\gamma,b}&=&-\lg T_{p}+\left(\lg I_{45}-2\lg P_{i,-3}-\lg
%f_b+52.3\right).\label{logLTP1}
%\end{eqnarray}
\begin{eqnarray}
\lg L_{\gamma,b,49}&=&-2\lg T_{p,3}+\left(0.6-2\lg B_{15}\right),\label{logLTB1}\\
\lg L_{\gamma,b,49}&=&-\lg T_{p,3}+\left(0.3-2\lg
P_{i,-3}\right),\label{logLTP1}
\end{eqnarray}
where a typical value of unit is adopted for $I_{45}$ and $R_6$ in
view of the sufficiently small variations of these parameters.
%Obviously, for fixed $B$ and
%$P_i$, the locations of the lines are only dependent on the free
%parameter $f_b$.
For a realistic magnetar, its rotation obviously cannot be rapider
than the Kepler rotation, at which the star starts shedding mass at
the equator. So an absolute upper limit on the spin periods of
magnetars is given by the Kepler period (Haensel et al. 2009)
\begin{eqnarray}
P_{K}=C\left(M\over1.4M_{\odot}\right)^{-1/2}R_6^{3/2}{\rm s},
\end{eqnarray}
where $M$ is the gravitational mass of the magnetar. For the
prefactor $C$, the careful numerical calculation given by Haensel et
al. (2009) showed that its value is slightly dependent on the
equation of state of the stellar matter, specifically, $C=0.783$ms
for NSs and $0.735$ms for SSs. For a conservative discussion, we
simply take $C=0.8$ms for both NSs and SSs as proposed by Lattimer
\& Prakash (2004).

In Fig. 1 we also scatter the observational internal-plateau data of
ten GRBs as listed in Table 1, which survived from an
internal-plateau test in Lyons et al. (2009) with three criterions:
(1) the X-ray light curve could not be adequately fitted by the
Willingale model (Willingale et al. 2007); (2) a relatively constant
X-ray flux lasts for a significant period of time; and (3) the
plateau is followed by a convincing steep decline with a temporal
index $>4$, so that the internal origin of the plateau is favored.
Under the isotropic assumption for the magnetar winds, the data
(open circles) is obviously in contradiction with the Kepler limit.
So as suggested by Lyons et al. (2009), we should adjust the
vertical locations of the data to fit the requirement of the Kepler
limit, by varying the value of the remaining free parameter
$\theta_w$. Then we get $\theta_w=45^\circ$ ($f_b= 0.3$) by assuming
the fastest possible period as the Kepler period. Taking this angle
as the beaming angle for each GRB, we replot the observational data
using solid asterisks in Fig. 1.

\section{{\it R}-mode limit on NSs and identifying SSs}
From Fig. 1, the magnetic field strengthes can be found to be around
typical values of $\sim 10^{15}-10^{16}$ Gauss for magnetars.
Besides that, we especially notice that a long-lasting rapid spin
with a period of $\sim 1-3$ms is required for the magnetars in GRBs
060607A and 070110. However, for a perfectly fluid compact star with
arbitrary rotation, {\r} oscillation can arise due to the action of
the Coriolis force with positive feedback and tends to be unstable
(Andersson 1998; Friedman \& Morsink 1998), succumbing to
gravitational radiation-driven Chandrasekhar-Friedman-Schutz
instability (Chandrasekhar 1970; Friedman \& Schutz 1978). As a
result, the rapid spin of newly born NSs can be reduced early and
remarkably by the strong gravitational radiation, but cannot for
nascent SSs.
%The couples between the gravitational
%radiation and the {\r}s are so strong in fact that the viscosities
%present in newly born NSs are not sufficient to suppress the
%instability. In contrast, nascent SSs would be not influenced by
%the {\r} instability
Then, the question arises as to whether the GRB rapidly-spinning
magnetars are NSs or SSs, or whether we can find SS candidates
from GRB magnetars.

Following a phenomenological model for {\r}s proposed by S\'a
(2004), the evolution of the amplitude of the {\r} oscillation,
$\alpha$, can be calculated from (Owen et al. 1998; S\'a \& Tom\'e
2005; Yu et al. 2009)
\begin{eqnarray}
{d\alpha\over
dt}&=&\left[1+{2\alpha^{2}\over15}({\delta}+2)\right]{\alpha\over\tau_g}-\left[1+{\alpha^{2}\over30}(4{\delta}+5)\right]
{\alpha\over\tau_v}+{\alpha\over2\tau_m},\label{alphat}
\end{eqnarray}
where ${\delta}$ is a free parameter describing the initial degree
of the differential rotation of a nascent magnetar,
\begin{eqnarray}
\tau_g=37{(P/P_K)^6}{\rm ~s}
\end{eqnarray}
is the gravitational radiation timescale, and
$\tau_{v}=(\tau_{sv}^{-1}+\tau_{bv}^{-1})^{-1}$ is the viscous
damping timescale with $\tau_{sv}$ and $\tau_{bv}$ corresponding
to the shear and bulk viscosities, respectively. Accordingly, the
spin-down of the magnetar against both gravitational and magnetic
braking effects is determined by
\begin{eqnarray}
{dP\over dt}={4\alpha^{2}\over15}({\delta}+2){P\over\tau_g}
-{\alpha^{2}\over15}(4{\delta}+5){P\over\tau_v}+{P\over\tau_m}.\label{omegat}
\end{eqnarray}
In the case of $\tau_{g}\gg\tau_m$ where the {\r}s cannot arise
sufficiently rapidly, the above equation can be well approximated by
Eq.(\ref{magbrake}). In contrast, for $\tau_{g}\ll\tau_m$, the
spin-down is instead dominated by the gravitational braking, if the
{\r} amplitude can reach its saturated value in time under a
condition of $\tau_{g}\ll\tau_{v}$. However, the value of $\tau_v$
is sensitive to the properties of the stellar matter.

To be specific, we display the viscous damping timescales for a NS
and a SS respectively as (Owen et al. 1998; Madsen
2000)\footnote{The viscous timescales of the SS presented here are
obtained for a normal quark matter phase. More generally, some color
superconducting phases such as the 2-flavor color superconductivity
and color-flavor-locked phases have also been suggested for
sufficiently high density and sufficiently low temperature (see a
recent review by Alford et al. 2008). In these phases, the
viscosities of the quark matter can be changed (generally reduced)
significantly. However, at high temperatures of
$\sim10^{10}-10^{11}$K as in nascent magnetars concerned in this
{\it letter}, the bulk viscosity of the superconducting matter could
be still comparable to that of the unpaired quark matter (e.g.,
Alford \& Schmitt 2007).}:
\begin{eqnarray}
\tau_{sv}^{NS}&\approx&2.52\times10^{10}{\mathcal{T}}_{10}^{8}~\rm s,\\
\tau_{bv}^{NS}&\approx&1.57\times10^{3}{\mathcal{T}}_{10}^{-6}(P/P_K)^2~\rm s;\\
\tau_{sv}^{SS}&\approx&2.49\times10^{10}{\mathcal{T}}_{10}^{5/3}~\rm s,\\
\tau_{bv}^{SS}&\approx&0.02{\mathcal{T}}_{10}^{-2}(P/P_K)^2~\rm s,
\end{eqnarray}
where ${\mathcal{T}}$ is the stellar temperature. For a nascent
magnetar with ${\mathcal{T}}= {\mathcal{T}}_i\sim10^{10}$K and
$P=P_i\sim P_K$, on one hand, the shear viscous damping of the {\r}s
can be neglected undoubtedly, no matter whether the star is a NS or
a SS due to $\tau_{sv}\gg(\tau_g,\tau_m,\tau_{bv})$. On the other
hand, the significant difference between $\tau_{bv}^{NS}$ ($\gg
\tau_g$) and $\tau_{bv}^{SS}$ ($\ll \tau_g$) would lead to two
distinct results: (1) For a nascent SS, the {\r} instability is
suppressed effectively by the bulk viscosity. By equaling $\tau_{g}$
to $\tau_{bv}^{SS}$ to get ${\mathcal{T}}_{10}\sim0.02$ and
following the direct-Urca-process-dominated cooling
${\mathcal{T}}={\mathcal{T}}_i(1+t/\tau_c)^{-1/4}$ with $\tau_c\sim
1$s, we can find that it should take about $\sim10^7$s to achieve
$\tau_{g}\sim\tau_{bv}^{SS}$ at which {\r}s can arise. During this
long period of time of $\sim10^7$s, the spin-down of the SS is
exclusively dominated by the relatively stronger magnetic dipole
radiation (Zheng et al. 2006). Therefore, the upper limit of the
spin periods of nascent SSs can be simply set at the Kepler period.
(2) In contrast, for nascent NSs, {\r}s can arise rapidly and arrive
at its saturated amplitude within hundreds of seconds (S\'a \&
Tom\'e 2005, 2006). As a result, the spin-down of the NSs would be
first dominated by the gravitational braking, much before the
magnetic braking operates (Yu et al. 2009).

Now let us investigate how long a nascent NS against {\r}
instability can keep constant spin with $\tau_{g}\ll\tau_{m}$. In
view of $\tau_{g}\ll(\tau_m,\tau_{v}^{NS})$, Eqs. (\ref{alphat}) and
(\ref{omegat}) can be solved analytically by ignoring the magnetic
and viscous terms. For convenience, we here exhibit an asymptotic
solution as (S\'a \& Tom\'e 2005, 2006)
%\begin{eqnarray}
%\alpha(t)&\approx&\left\{
%\begin{array}{ll}
%\alpha_i\exp\left({t\over\tau_{g,i}}\right),&{\rm for}~t<T_{g}\\
%{3.56\over\sqrt{{\delta}+2}}\left({t\over\tau_{g,i}}\right)^{1/10},&{\rm
%for}~t>T_{g}
%\end{array}\right.\label{alphatan}
%\end{eqnarray}
\begin{eqnarray}
P(t)&\approx&\left\{
\begin{array}{ll}
P_i\left[1-{2\alpha_i^2\over15}({\delta}+2)\exp\left({2t\over\tau_{g,i}}\right)\right]^{-1},&{\rm for}~t<T_{g}\\
1.6P_i\left({t\over\tau_{g,i}}\right)^{1/5},&{\rm for}~t>T_{g}
\end{array}\right.
\end{eqnarray}
where $\alpha_i$ is the initial \textit{r}-mode amplitude. The
break time $T_{g}$ can be solved from $({d^2\alpha/
dt^2})_{t=T_{g}}=0$ to be
\begin{eqnarray}
T_{g}&=&-37\left[\ln\alpha_i+{1\over2}\ln({\delta}+2)+{1\over2}\ln{6\over5}\right]\left({P_i\over
P_K}\right)^6\nonumber\\
&\equiv&T_{K}(\alpha_i,{\delta})\left({P_i\over P_K}\right)^6,
\end{eqnarray}
%At this time, the spin period reads
%\begin{eqnarray}
%P(T_{g})\approx
%P_i\left[1+{4\over3}({\delta}+2)Q\alpha_{\max}^2\right]^{-1}=1.125P_i.
%\end{eqnarray}
at which $P=1.125P_i$. After $T_{g}$, the spin period of the NS can
no longer be regarded as a constant and the luminosity of the wind
evolves as $L_{w}\approx L_{msd}\propto t^{-4/5}$, which obviously
deviates from a plateau emission. So we conclude that {\it a nascent
NS with an initial period $P_i$ can keep nearly constant spin for at
most a period of time of $T_{g}(P_i)$.} For $P_i=P_K$, we have
$T_{g}=T_K$. Within a wide parameter region of
$10^{-10}<\alpha_i<10^{-6}$ and $0<{\delta}<10^{8}$ (S\'a \& Tom\'e
2005, 2006), the value of $T_K$ varies from 170 s to 840 s.

\begin{figure}
%\centerline{\psfig{figure=f2.eps,width=15cm}}
\plotone{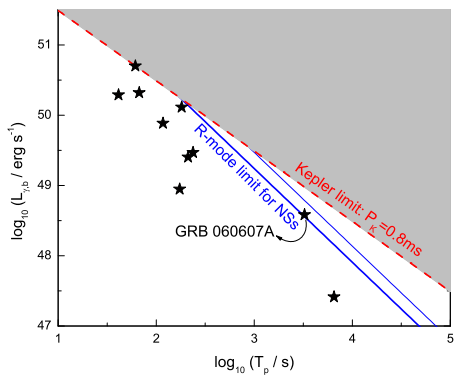}\caption{The {\r} limit on the spin of NSs for
$T_K=170$ s (thick solid line) and $T_K=680$ s (thin solid line).
The Kepler limit is represented by the thick dashed line, above
which the shaded region is forbidden for magnetars. The beaming
angle is taken to be $\theta_{w}=45^{\circ}$ for the observational
data in order to make the fastest period just longer than the Kepler
period.} \label{figure:2}
\end{figure}
Therefore, for GRB magnetars with $T_p>10^3$s, it is necessary to
test whether their spin is sufficiently slow to determine a
sufficiently long $T_{g}$ as
\begin{eqnarray}
P_i<P_{R}=P_K\left(T_p/T_K\right)^{1/6}\label{PR}.
\end{eqnarray}
If not, the magnetar is quiet unlikely to be a NS. Alternatively,
it could be a SS candidate. Substituting Eq.(\ref{PR}) into
(\ref{logLTP1}) with $P_{K,-3}=0.8$, we can obtain
\begin{eqnarray}
\lg L_{\gamma,b,49}&=&-{4\over3}\lg
T_{p,3}+\left(-0.5+{1\over3}\lg T_K\right).\label{logLT2}
\end{eqnarray}
With a certain value of the free parameter $T_{K}$, this equation
determines a {\r} limit line for NSs in the $\lg L_{\gamma,b}-\lg
T_{p}$ panel, as shown by the solid lines in Fig. 2, above which
$T_p>T_{g}$.

{\it Then an especial region below the Kepler limit line and above
the {\r} limit line is highlighted in the $\lg L_{\gamma,b}-\lg
T_{p}$ panel, where only SSs can appear.} When we use this SS-only
region to cover the observational GRB internal-plateau data, we can
obtain that either (1) some SS candidates are indicated by the data
locating in the SS-only region, or (2) all data are clearly below
the {\r} limit line. In the latter case, the SS model cannot be in
principle ruled out, but the influence of the gravitational
radiation-driven {\r} instability on the spin of NSs would be
confirmed. For the small sample of GRB magnetars listed in Table 1,
we find in Fig. 2 that GRB 060607A locates just above the {\r} limit
line, by assuming the fastest possible period as the Kepler period
($\theta_{w}=45^{\circ}$) and taking $T_K=170$s. This implies that
the magnetar in GRB 060607A could be a SS. It is fair to say,
though, that the uncertainties and approximations involved could be
large enough that the NS model might be still available for this
magnetar. So a much richer GRB sample is sorely demanded in order to
get a more definite conclusion.

\section{Summary and Discussion}
The internal-plateau emission is considered to indicate that a
central rapidly-spinning magnetar is formed during some GRBs. Such a
newly born magnetar is sometimes required to be able to keep a rapid
spin with a sub- or millisecond period for thousands of seconds.
Then, according to the spin limit of nascent NSs due to {\r}
instability, we propose a method to identify SS candidates from the
GRB magnetars, since the {\r} instability cannot operate in nascent
SSs. Despite the smallness of the present GRB magnetar sample, the
effectiveness of the method is somewhat exhibited by the appearance
of GRB 060607A.

A few decades ago, the concept of SS was proposed formerly as a new
and novel class of compact stars (Alcock et al. 1986; Haensel et al.
1986) and even as GRB central objects (Cheng \& Dai 1996, 1998; Dai
\& Lu 1998a), based on a conjecture that the true ground state of
matter is strange quark matter rather than Fe$^{56}$ (Witten 1984).
In both physics and astronomy, it is of significant importance to
identify or rule out SSs from observational astrophysical objects.
Seasonably, a plausible method was designed by Madsen (1998)
according to the distinct rotational properties of NSs and SSs
against the {\r} instability. However, for most detected Galactic
millisecond pulsars, the stellar temperature is generally much lower
than $\sim10^{10}$K, in which case some complications arise from the
shear viscosity due to the stellar crust (e.g., Bildsten \&
Ushomirsky 2000) and the superfluidity appearing in the stellar
core. This makes it difficult to distinguish SSs from NSs within the
Galactic pulsar sample (Madsen 2000). In contrast, the situation in
newly born pulsars would be much clearer due to their sufficiently
high temperature. Moreover, for GRB magnetars, a relatively richer
sample could be selected from the huge GRB storage, in contrast to
the rare observation of newly born pulsars in the Galaxy. In view of
these advantages, it is worthwhile to use the newly-born GRB
magnetars to test the SS hypothesis.

However, it should be notice that the uncertainties and
approximations of the model still prohibit the present small GRB
magnetar sample from becoming a gold sample for testing SS
hypothesis. Therefore, it is demanded to make great efforts in
following aspects: (1) a thorough investigation on magnetar wind.
Here we would like to point out that the angle of the magnetar
wind is probably different from that of the GRB ejecta, in respect
that the GRB ejecta could be driven by an hyperaccretion onto the
magnetar rather than the magnetar wind (Zhang \& Dai 2008, 2009).
(2) finding other constraints on the wind angle. For example, in
order to build up sufficient dynamo action responsible for the
intense magnetic fields, the initial spin periods of magentars
could be required to be $\leq10$ ms (Lyons et al. 2009; Usov
1992), which leads to $\theta_{w}>16^{\circ}$; (3) accumulating a
much larger number of GRB magnetars.
%This phenomena is generally considered to mark the most violent,
%cataclysmic explosions in the Universe, such as collapse of
%massive stars and coalescence of compact binaries, although there
%is still a long way to go to confirm the nature of the GRB
%progenitors. although the GRB central engines and their
%progenitors are still open questions. Thank to the {\it Swift}, a
%chance for ``directly" viewing the central objects is provided by
%the internal plateau afterglow emission

 \acknowledgements We would like to acknowledge Z.G. Dai, K.S. Cheng, and Xue-Wen Liu for their useful suggestions.
This work is supported by the National Natural Science Foundation of
China (grant no. 10773004).

\end{document}